\documentclass[12pt,a4paper]{article}

\newcommand{\la}{\lambda}
\newcommand{\me}{\mathcal{E}}
\newcommand{\ms}{\mathcal{S}}

\begin{document}

\begin{flushright}
hep-th/0403180
\end{flushright}
\vspace{1.8cm}

\begin{center}
 \textbf{\Large Folded Three-Spin String Solutions in $AdS_5 \times S^5$}
\end{center}
\vspace{1.6cm}
\begin{center}
 Shijong Ryang
\end{center}

\begin{center}
\textit{Department of Physics \\ Kyoto Prefectural University of Medicine
\\ Taishogun, Kyoto 603-8334 Japan}  \par
\texttt{ryang@koto.kpu-m.ac.jp}
\end{center}
\vspace{2.8cm}
\begin{abstract}
We construct a spinning closed string solution in $AdS_5 \times S^5$
which is folded in the radial direction and has two equal spins in
$AdS_5$ and a spin in $S^5$. The energy expression of the three-spin
solution specified by the folding and 
winding numbers for the small $S^5$ spin
shows a logarithmic behavior and a one-third power behavior of the
large total $AdS_5$ spin, in the long string and in the short string
located near the boundary of $AdS_5$ respectively.
It exhibits the non-regular expansion in the 't Hooft coupling constant,
while it takes the regular one when the $S^5$ spin becomes large.
\end{abstract} 
\vspace{3cm}
\begin{flushleft}
April, 2004
\end{flushleft}

\newpage
\section{Introduction}

The AdS/CFT correspondence \cite{JM,GKP,EW} has been studied in the 
supergravity approximation. In order to verify the 
correspondence in its full extent
it is desirable to go beyond the low energy supergravity approximation.
Under the circumstance that it is hard to quantize the superstring 
theory in $AdS_5 \times S^5$, the solvability of the string theory 
in the pp-wave background  \cite{RM,MT} has opened a new step and led to
an interesting proposal identifying particular stringy oscillator
states with gauge invariant near-BPS operators with large R-charge
in the BMN limit for the $\mathcal{N}=4$ SU(N) super Yang-Mills (SYM)
theory \cite{BMN}. This proposal has been interpreted as a special case of
a semiclassical expansion of $AdS_5 \times S^5$ string theory selecting
a particular sector of states where the string shrinks to a point
and moves with a large angular momentum along a large circle of 
$S^5$ \cite{GK} and further developed 
\cite{FT,JR,ABP,MSW,AG,JAM,AT,AB,AM}.
A considerable amount of work has followed for
various semiclassical string or membrane solutions 
\cite{HN,RV,LO,PT,MAM,SR,DFR,LL,APB,DAB,MS,PBV,KL}.

Motivated by attempts to AdS/CFT correspondence to non-BPS states there
have been the constructions of semiclassical string solutions with 
several angular momenta in different directions of $AdS_5$ and $S^5$.
A novel solution describing a circular closed string rotating 
simultaneously in two planes in $AdS_5$ with equal spins $S$ has been
shown to produce a large-spin  expansion of the energy whose subleading
term scales with spin as $S^{1/3}$ instead of $\log S$ in the one-spin
solution, when the point-like string is located close to the 
boundary of $AdS_5$ \cite{SFT}. There have been the other solutions 
describing a circular closed string with three SO(6) angular momenta
in $S^5$ \cite{SFT,FAT} and a folded closed string with two SO(6) angular
momenta \cite{FTA} located at the origin of $AdS_5$. A folded
closed string solution with three spins $(J_1,J_2,J_3)$ in $S^5$
has been constructed as a periodic solution of a Neumann 
one-dimensional integrable system \cite{AFR}, where 
the string is stretched along two angular directions 
and bent at one point. A three-spin string solution
of circular type has been also analyzed and a two-spin one has
been explicitly represented by the elliptic functions and a winding 
number. Their space-time energies expressed by the spins and the string
tension $\sqrt{\la}$ are arranged to take the regular expansions in 
the 't Hooft coupling constant $\la$ and remarkably matched onto the
associated scaling dimensions of the non-BPS operators obtained by 
analyzing the relevant Bethe equation for a spin chain model in the 
perturbative SYM theory \cite{MZ,BMS,FTA,BS,AFR,BFS,JE,AAT}.

A hybrid $(S,J)$ state represented by a string rotating in one
plane in $S^5$ and having also one large spin in $AdS_5$ has been shown to
be analytically continued to the $(J_1,J_2)$ state represented by a folded
string rotating in two planes in $S^5$ \cite{BFS}. A more general solution
describing a string rotating in both $AdS_5$ and $S^5$ with constant radii
has been constructed as a periodic solution of a Neumann-Rosochatius
one-dimensional integrable system to be 
characterized by the $2+3$ spins $(S_a,J_i)$
and the $2+3$ winding numbers \cite{ART}. Its space-time energy has a 
regular large-spin expansion if there is at least one large spin in $S^5$.
There have been various investigations about the multi-spin string
solutions and the gauge-string duality \cite{BGK,DNW,AGY,MMT,GAS,
AMV,MK,BJS,NK,ALK,AMY,CK}. In ref. \cite{ALK} 
a spinning solution folded in 
the radial direction with two equal spins in $AdS_5$ has been presented,
where the string is orbiting around the origin and the radius of string
is not constant but oscillates between two locations. This solution 
specified by the folding and winding numbers 
associated with the radial direction and
an angular direction of $S^3$ within $AdS_5$ shows a logarithmic behavior
of the energy-spin relation for the long string.

We will construct a folded three-spin solution which has not only the two
equal spins in $AdS_5$ but also a spin in $S^5$ and is stretched along
both the radial and angular directions. It will be shown how this hybrid
three-spin solution interpolates between the two-spin solution with the
logarithmic energy-spin relation for the long string
and that with the $S^{1/3}$ relation for the point-like string.

\section{Folded three-spin solutions}

The bosonic part of the type IIB superstring action for the $AdS_5 
\times S^5$ background is given by 
\begin{equation}
I = -\frac{\sqrt{\la}}{4\pi}\int d^2\xi [ G_{mn}^{(AdS_5)}(x)
\partial_ax^m\partial^ax^n + G_{pq}^{(S^5)}(y) \partial_ay^p
\partial^ay^q ], \hspace{1cm} \sqrt{\la} \equiv \frac{R^2}{\alpha'},
\end{equation}
where the $AdS_5 \times S^5$ metric in the global coordinates is 
expressed as 
\begin{eqnarray}
(ds^2)_{AdS_5} &=& d\rho^2 - \cosh^2\rho dt^2 + \sinh^2\rho (d\theta^2 +
\cos^2\theta d\phi_1^2 + \sin^2\theta d\phi_2^2), \nonumber \\
(ds^2)_{S^5} &=& d\gamma^2 + \cos^2\gamma d\varphi_3^2 + \sin^2\gamma
(d\psi^2 + \cos^2\psi d\varphi_1^2 + \sin^2\psi d\varphi_2^2).
\end{eqnarray}
We consider a configuration that a closed string is spinning in the 
$\phi_1$ and $\phi_2$ directions of $AdS_5$ with equal spin 
as well as in the $\varphi_3$ direction of $S^5$, and stretched along
the radial coordinate $\rho$ and along the angular coordinate $\theta$ of
$S^5$. We make the following ansatz describing this configuration 
\begin{eqnarray}
&t=\kappa\tau, \hspace{0.5cm} \phi_1 = \omega \tau,  
\hspace{0.5cm}\phi_2 = \omega \tau, \hspace{0.5cm}
\varphi_3 = \nu \tau, \nonumber \\
&\rho = \rho(\sigma) = \rho(\sigma+2\pi), \hspace{1cm} \theta = 
\theta(\sigma) = \theta(\sigma+2\pi), \\
&\gamma = \psi = \varphi_1 = \varphi_2 = 0. \nonumber
\end{eqnarray}
The string equations of motion for $\theta$ and $\rho$ are given by
\begin{equation}
(\sinh^2\rho \theta')' = 0, \hspace{1cm} \rho'' - \sinh\rho \cosh\rho
(\kappa^2 + \theta'^2 - \omega^2 ) = 0,
\end{equation}
which lead to 
\begin{eqnarray}
\theta' = \frac{c}{\sinh^2\rho}, \hspace{1cm} c = \mathrm{const}, 
\label{thc} \\
\rho'' - c^2\frac{\cosh\rho}{\sinh^3\rho} - \frac{1}{2}(\kappa^2 -
\omega^2)\sinh 2\rho = 0.
\label{rhc}\end{eqnarray}
The conformal gauge constraint is described by
\begin{equation}
\rho'^2 = \kappa^2\cosh^2\rho - \frac{c^2}{\sinh^2\rho} - 
\omega^2\sinh^2\rho - \nu^2,
\end{equation}
which is also the first integral of (\ref{rhc}) with an appropriate 
integral constant. In terms of $x = \cosh\rho$ it can be rewritten by
\begin{equation}
x'^2 = (\omega^2 - \kappa^2)(a_+ - x^2)(x^2 - a_-),
\label{xoa}\end{equation}
where the constants $a_{\pm}$ are
\begin{equation}
a_{\pm} = \frac{1}{2(\omega^2 - \kappa^2)}[2\omega^2 - \kappa^2 - \nu^2
\pm \sqrt{\kappa^4 + 2(2c^2-\nu^2)\kappa^2 + \nu^4 - 4\omega^2 c^2} ].
\end{equation}
We impose a condition 
\begin{equation}
1 \le a_- < a_+
\label{apm}\end{equation}
so that the string has two different turning points $a_+, a_-$ in the 
radial direction. The integral of motion $c$ and the parameter $\kappa$ 
that is associated with the energy of the system, are described in terms
of the turning points $a_+, a_-$ as
\begin{eqnarray}
c^2 &=& (\omega^2 - \kappa^2)(a_+ - 1)(a_- - 1),  \\
\kappa^2 &=& \omega^2 - \frac{\omega^2 - \nu^2}{a_+ + a_- -1},
\label{kao}\end{eqnarray}
which imply that the condition (\ref{apm}) yields $\nu < \kappa < \omega$.
Therefore it follows that $a_- = 1$ for $c = 0$.

We choose a differential equation  $x' = -[(\omega^2 - \kappa^2)
(a_+ - x^2)(x^2 - a_-)]^{1/2}$ for (\ref{xoa}), which is transformed
through the change of variables $x/\sqrt{a_+} = y, 
\sqrt{a_+(\omega^2-\kappa^2)} \sigma = u$ into
\begin{equation}
\frac{d}{du}y(u) = - \sqrt{(1 - y^2)\left( y^2 - \frac{a_-}{a_+}\right)}.
\label{uya}\end{equation}
This expression yields a solution expressed by the Jacobi elliptic 
function
\begin{eqnarray}
\cosh\rho &=& \sqrt{a_+} \mathrm{dn}(\sqrt{a_+(\omega^2-\kappa^2)}
\sigma,m), \nonumber \\
 &=& \left(a_+ - (a_+ - a_-)\mathrm{sn}^2(\sqrt{a_+(\omega^2-\kappa^2)}
\sigma,m)\right)^{1/2}, \hspace{0.5cm} m = \frac{ a_+ - a_-}{a_+}.
\label{cdn}\end{eqnarray}
When $a_{\pm}$ are parametrized as $a_{\pm} = \cosh^2\rho_{\pm}$, 
from this expression we see
that $\rho$ indeed oscillates between $\rho_+$ and $\rho_-$.
The periodicity condition $\rho(\sigma) = \rho(\sigma + 2\pi)$ is 
expressed in terms of a folding number $N$ as
\begin{equation}
\frac{2\pi}{N}\sqrt{a_+(\omega^2-\kappa^2)} = 2K(m)
\label{ank}\end{equation}
since dn($u$) and $\mathrm{sn}^2(u)$ in (\ref{cdn}) have a fundamental
period $2K$ where $K$ is the complete elliptic integral of the first kind.
In this string configuration composed of 
$2N$ segments $\rho$ starts at $\rho_+$
and becomes $\rho_-$ as $\sigma$ goes from 0 to $\pi/N$ for one segment.
Alternatively the periodicity condition (\ref{ank}) can be derived by
the direct integration of (\ref{xoa}) as
\begin{equation}
\int_0^{2\pi}d\sigma = 2N\int_{\sqrt{a_+}}^{\sqrt{a_-}}dx 
\frac{-1}{\sqrt{(\omega^2 - \kappa^2)(a_+ - x^2)(x^2 - a_-)}}.
\label{inn}\end{equation}
Substituting the solution (\ref{cdn}) into (\ref{thc}) and integrating 
we have 
\begin{equation}
\theta(\sigma) = \sqrt{\frac{a_- - 1}{a_+(a_+ -1)}}\Pi \left(
\sqrt{a_+(\omega^2-\kappa^2)}\sigma, \frac{a_+m}{a_+ - 1}, m \right), 
\hspace{1cm} m = \frac{a_+ - a_-}{a_+},
\label{thp}\end{equation}
where $\Pi(u,n,m)$ is the elliptic integral of the third kind and we have
chosen an integration constant such that $\theta(0) = 0$.
From the periodicity condition $\theta(2\pi) = \theta(0) + 2\pi M$ for the
angular coordinate the winding number $M$ is specified as
\begin{equation}
2\pi M = N\theta\left(\frac{2\pi}{N}\right) = 
2N\theta\left(\frac{\pi}{N}\right),
\end{equation}
which combines with (\ref{thp}) and (\ref{ank}) to yield 
\begin{equation}
\pi\frac{M}{N} = \sqrt{\frac{a_- - 1}{a_+(a_+ -1)}}\Pi\left(
\frac{a_+ - a_-}{a_+ - 1}, m \right),
\label{pap}\end{equation}
where $\Pi(n,m) \equiv \Pi(K,n,m)$. 
Alternatively it follows from the direct
integration of (\ref{thc}) together with (\ref{xoa}) 
\begin{equation}
\int_0^{2\pi M}d\theta = 2N \int_{\sqrt{a_+}}^{\sqrt{a_-}}dx
\frac{-1}{\sqrt{(\omega^2 - \kappa^2)(a_+ - x^2)(x^2 - a_-)}}
\frac{c}{x^2 - 1}.
\end{equation}
The integers $N, M$ will label different spinning configurations.

\section{Energy-spin relations}

The energy and three spins of spinning string solution are given by
\begin{eqnarray}
E &=& \sqrt{\la}\kappa\int_0^{2\pi}\frac{d\sigma}{2\pi} \cosh^2\rho
 \equiv \sqrt{\la} \me, \hspace{1cm} J = \sqrt{\la}\nu, \nonumber \\
S_1 &=& \sqrt{\la}\omega\int_0^{2\pi}\frac{d\sigma}{2\pi} \sinh^2\rho 
\cos^2\theta  \equiv \sqrt{\la} \ms_1, \\
S_2 &=& \sqrt{\la}\omega\int_0^{2\pi}\frac{d\sigma}{2\pi} \sinh^2\rho 
\sin^2\theta  \equiv \sqrt{\la} \ms_2, \nonumber
\end{eqnarray}
where $J, S_1$ and $S_2$ are the spins coming from the rotations in the
$\varphi_3, \phi_1$ and $\phi_2$ directions, respectively.
There is a relation between them
\begin{equation}
\frac{\me}{\kappa} - \frac{\ms}{\omega} = 1, \hspace{1cm}
\ms \equiv \ms_1 + \ms_2.
\label{ess}\end{equation}
In view of (\ref{inn}) the energy is computed as
\begin{equation}
\me = 2N\kappa\int_{\sqrt{a_+}}^{\sqrt{a_-}}\frac{dx}{2\pi}\frac{-x^2}
{\sqrt{(\omega^2 - \kappa^2)(a_+ - x^2)(x^2 - a_-)}}
= \frac{N\kappa \sqrt{a_+}}{\pi\sqrt{\omega^2 - \kappa^2}}E(m),
\label{ein}\end{equation}
where $E(m)$ is the complete elliptic integral of the second kind.
The substitution of $\kappa$ in (\ref{kao}) into (\ref{ein}) provides
\begin{equation}
\me = \frac{N\sqrt{a_+}}{\pi}\left( \frac{\nu^2 + \omega^2(a_+ + a_- - 2)}
{\omega^2 - \nu^2} \right)^{1/2} E(m).
\label{ena}\end{equation}
From (\ref{ank}) and (\ref{kao}), the angular velocity parameter $\omega$
is given by
\begin{equation}
\omega = \left( \nu^2 + \frac{a_+ + a_- - 1}{\pi^2a_+}K(m)^2N^2   
\right)^{1/2}
\label{okn}\end{equation}
so that the energy expression (\ref{ena}) is expressed as 
\begin{equation}
\me = \frac{a_+E}{K}\left( \nu^2 + \frac{a_+ + a_- - 2}{\pi^2a_+}K^2N^2
\right)^{1/2}.
\label{nue}\end{equation}
On the other hand we use (\ref{okn}) and (\ref{kao}) to rewrite the 
relation (\ref{ess}) as
\begin{equation}
\me = \left( 1 + \frac{\ms}{\sqrt{\nu^2 + \frac{a_+ + a_- - 1}{\pi^2a_+}
K^2N^2}} \right)\left( \nu^2 + \frac{a_+ + a_- - 2}{\pi^2a_+}K^2N^2
\right)^{1/2}.
\label{esn}\end{equation}
The comparison between (\ref{nue}) and (\ref{esn}) yields
the total $AdS_5$ spin
\begin{equation}
\ms = \left( \frac{a_+E}{K} - 1 \right)\left( \nu^2 + 
\frac{a_+ + a_- - 1}{\pi^2a_+}K^2N^2\right)^{1/2}.
\label{ska}\end{equation}
In order to analyze the above equations it is convenient to use two 
parameters $m = (a_+ - a_-)/a_+, n = (a_+ - a_-)/(a_+ - 1)$ instead of the
turning points $a_+, a_-$. Thus from (\ref{nue}), (\ref{ska}) and 
(\ref{pap}) we have the following three equations
\begin{eqnarray}
\me &=& \frac{nE(m)}{n-m} \left( \frac{\nu^2}{K(m)^2} + 
\frac{m(2-n)}{\pi^2 n} N^2 \right)^{1/2}, \label{ekm} \\
\ms &=& \left( \frac{nE(m)}{n-m} - K(m) \right) \left( \frac{\nu^2}
{K(m)^2}+ \frac{n + m - nm}{\pi^2 n} N^2 \right)^{1/2}, \label{skm} \\
\pi \frac{M}{N} &=& \sqrt{\frac{(1-n)(n-m)}{n}} \Pi(n, m),
\label{mnp}\end{eqnarray}
which are expressed in terms of the auxiliary two independent parameters
$m, n$ so that if $m, n$ are eliminated from the three transcendental
equations, the energy $E$ is in principle derived as a function of
$S, J, M$ and $N$. The relation (\ref{esn}) is also expressed
in terms of $m, n$ as
\begin{equation}
\me = \left( 1 + \frac{\ms}{\sqrt{\nu^2 + \frac{n + m - nm}{\pi^2 n}
K(m)^2N^2}}\right)
\left( \nu^2 +  \frac{m(2-n)}{\pi^2 n}K(m)^2 N^2 \right)^{1/2}.
\label{rel}\end{equation}

Now we are ready to extract explicit analytic expressions of energy-spin
relations in several special regions 
specified by $J$ or $S \gg \sqrt{\la}$
and  $J$ or $S \ll \sqrt{\la}$. The condition (\ref{apm}) is translated 
into $0 < m < n \le 1$. The turning points $\rho_+, \rho_-$ are written
in terms of $m, n$ as 
\begin{equation}
\cosh^2\rho_+ = \frac{n}{n-m}, \hspace{1cm} \cosh^2\rho_- = 
\frac{n(1-m)}{n-m}. 
\end{equation}
In view of $ \cosh^2\rho_+ - \cosh^2\rho_- = m/(1-\frac{m}{n})$, 
a parameter region $n \approx m \approx 1$ corresponds 
to the long string, while a region
specified by $m/n \ll 1$ and $m \ll 1$ to the short string.
There is the other interesting short string region 
$n \approx m \ll 1- \frac{m}{n} \ll 1$.

We start to consider the long string case and write down a formula
\begin{eqnarray}
\Pi(n,m) &=& K(m) + \frac{2\sqrt{1-k'^2\sin^2\psi}}{k'^2\sin 2\psi}
\Bigg[F(\psi, k'^2)K(m) - E(\psi, k'^2)K(m) \nonumber \\
 &-& F(\psi, k'^2)E(m) + \frac{\pi}{2} \Bigg],
\end{eqnarray}
where $k'^2 = 1-m, n= 1- k'^2 \sin^2\psi$, and $F(\psi, k'^2)$ and 
$E(\psi, k'^2)$ are the elliptic integrals of the first and second kind
respectively. For $n \approx m \approx 1$ it approximately reduces to  
\begin{equation}
\Pi(n, m) \approx \sqrt{\frac{n}{(1-n)(n-m)}}\left( \frac{\pi}{2} -
\sin^{-1} \sqrt{\frac{1-n}{1-m}} \right),
\end{equation}
whose substitution into (\ref{mnp}) provides
\begin{equation}
\sqrt{\frac{1-n}{1-m}} \approx \cos \frac{M}{N} \pi.
\label{com}\end{equation}
We can now use (\ref{com}) to rewrite (\ref{ekm}) and (\ref{skm}) as
\begin{eqnarray}
\me &\approx& \left( \frac{\nu^2}{K^2} + \frac{N^2}{\pi^2} \right)^{1/2}
\frac{1}{(1-m)\sin^2\left(\frac{M}{N}\pi\right)} 
\left( 1 + \frac{1-m}{2} \log \frac{16}{1-m} \right), \\
\ms &\approx& \left( \frac{\nu^2}{K^2} + \frac{N^2}{\pi^2} \right)^{1/2}
\frac{1}{(1-m)\sin^2\left(\frac{M}{N}\pi\right)} 
\left[ 1 + \frac{1-m}{2} \left(1- \sin^2\left(
\frac{M}{N}\pi\right)\right)\log \frac{16}{1-m} \right]
\end{eqnarray}
with $K(m) \approx \frac{1}{2}\log\frac{16}{1-m}$. 
Their difference gives
\begin{equation}
\me - \ms \approx \frac{1}{2}\left( \frac{\nu^2}{K^2} + 
\frac{N^2}{\pi^2} \right)^{1/2}\log\frac{16}{1-m}.
\end{equation}

For the region $\nu/\log(\frac{1}{1-m}) \ll 1$, the large total spin $\ms$
is approximately estimated as
\begin{equation}
\ms \approx \frac{N/\pi}{(1-m)\sin^2\left(\frac{M}{N}\pi\right)}
\end{equation}
so that we have a logarithmic scaling behavior
\begin{equation}
\me - \ms \approx \frac{N}{2\pi}\log \pi\ms + \frac{\pi\nu^2}{N\log\pi\ms}
+ \cdots.
\end{equation}
Restoring the $\la$-dependence provides
\begin{equation}
E - S \approx \frac{N\sqrt{\la}}{2\pi} \log\frac{\pi S}{\sqrt{\la}} +
\frac{\pi J^2}{N\sqrt{\la}\log\frac{\pi S}{\sqrt{\la}}} 
+ \cdots,  \hspace{1cm}
\frac{S}{\sqrt{\la}}\gg 1, \; \frac{J}{\sqrt{\la}}\ll \log
\frac{S}{\sqrt{\la}},
\label{loe}\end{equation} 
which shows a semiclassical $\sqrt{\la}$ 
expansion for the enegry-spin relation.
When $J = 0$ this expression reduces to the result of ref. \cite{ALK}.

For the opposite region $\nu/\log(\frac{1}{1-m}) \gg 1$, we have an 
approximate relation $\log(\ms/\nu)\approx \log(1/(1-m))$, which yields
\begin{equation}
\me - \ms \approx \nu + \frac{N^2}{8\pi^2\nu} \log^2\frac{\ms}{\nu}
+ \cdots.
\end{equation}
Thus we have a regular expansion in integer powers of $\la$
\begin{equation}
E - S \approx J + \frac{\la N^2}{8\pi^2 J} \log^2\frac{S}{J}
+ \cdots, \hspace{1cm} \frac{S}{\sqrt{\la}} \gg 1, 
\; \frac{J}{\sqrt{\la}} \gg \log\frac{S}{J}.
\label{esj}\end{equation}

Now we analyze the short string region specified by $m \ll n \ll 1$
that is more restricted than $m \ll n < 1$. Eq. (\ref{mnp}) 
is approximately given by
\begin{equation}
\frac{2M}{N} \approx \sqrt{1-\frac{m}{n}},
\end{equation}
which is satisfied when $2M < N$. The radial location of 
short string is so specified by 
\begin{equation}
\cosh^2\rho_+ \approx \cosh^2\rho_- \approx \left(\frac{N}{2M}\right)^2
\end{equation}
that the string is located near the origin of $AdS_5$.

For $\nu \ll 1$ from (\ref{skm})  the total spin 
$\ms$ is approximately estimated to be a small value
\begin{equation}
\ms \approx \frac{N}{2} \frac{m}{n} \ll 1
\end{equation}
and then the energy $\me$ in (\ref{ekm}) is specified by
\begin{equation}
\me^2 \approx \nu^2 + \ms N.
\end{equation}
This expression becomes
\begin{equation}
E^2 \approx J^2 + \sqrt{\la}N S, \hspace{1cm} \frac{S}{\sqrt{\la}}\ll 1,
\; \frac{J}{\sqrt{\la}} \ll 1,
\label{reg}\end{equation}
which shows the usual Regge trajectory relation when $J = 0$.
The energy-spin relations of three-spin $(S_1=S_2, J)$ solution 
(\ref{loe}), (\ref{esj}) and (\ref{reg}) for the respective parameter 
regions are similar to the analogous relations of the two-spin $(S, J)$
solution \cite{FT,JR}.

For $\nu \gg 1$ eq. (\ref{skm}) can be approximately expressed as
\begin{equation}
\ms \approx \frac{m}{n-m} \left(\nu^2 + \frac{N^2}{4}
\left( 1+ \frac{m}{n} \right) \right)^{1/2}
\label{mns}\end{equation}
from which it is noted that $\ms \ll \nu$. 
Taking advantage of $\ms \ll \nu$ to use an iterative expansion method
we can derive a solution of (\ref{mns}) in a large-spin expansion form as
\begin{equation}
\frac{m}{n} \approx \frac{\ms}{\nu} - \left(\frac{\ms}{\nu}\right)^2
- \frac{N^2\ms}{8\nu^3} + \frac{N^2\ms^2}{8\nu^4} + \cdots.
\label{exs}\end{equation}
Substituting this solution into the energy expression (\ref{rel}) 
instead of (\ref{ekm}) we have 
\begin{equation}
\me \approx \nu + \ms + \frac{N^2\ms}{8\nu^2} - \frac{N^2\ms^2}{8\nu^3}
+ \cdots,
\end{equation}
which reads
\begin{equation}
E \approx J + S + \frac{\la N^2S}{8J^2} - \frac{\la N^2S^2}{8J^3}
+ \cdots, \hspace{1cm} \frac{J}{\sqrt{\la}} \gg 1, \; J \gg S.
\label{jse}\end{equation}
The energy expression (\ref{jse}) for the three-spin solution in a near
BMN limit $S/J \ll 1$ is compared with that for the two-spin $(S, J)$ 
solution \cite{FT,JR,BFS}. Here we write down the energy of the two-spin
solution including full dependence on $\la$ presented in \cite{BFS}
\begin{equation}
E = J + S\sqrt{1+ \frac{\la}{J^2}} - \frac{\la S^2}{4J^3} 
\frac{2 + \la/J^2}{1 + \la/J^2} + \cdots,
\end{equation}
whose expansion in $\la/J^2$ 
\begin{equation}
E \approx J + S + \frac{\la S}{2J^2} - \frac{\la S^2}{2J^3} + \cdots 
\end{equation}
has a similar behavior to (\ref{jse}). The analytic expansion in $\la$ of
(\ref{jse}) is also compared with the non-analytic one of the energy 
for (\ref{reg}).

Let us consider the parameter region $n \approx m \ll 1- \frac{m}{n}
\ll 1$ where the string becomes short and its radial location takes
a large value $\cosh^2\rho_+^2 \approx \cosh^2\rho_-^2 
\approx 1/(1- \frac{m}{n})$ so that the short string 
is close to the boundary of $AdS_5$. In this region the ratio $M/N$ is
specified by
\begin{equation}
\frac{2M}{N} \approx \sqrt{1 - \frac{m}{n}} \ll 1.
\label{rat}\end{equation}

For $\nu \ll 1$ eq. (\ref{skm}) is also expressed as (\ref{mns}), from 
which it is noted that $\ms \gg 1$. Therefore the solution of (\ref{mns})
is approximately obtained by the iterative expansion procedure as
\begin{equation}
\frac{m}{n} \approx 1 - \frac{N}{\sqrt{2}\ms} + \frac{5N^2}{8\ms^2}
- \frac{\nu^2}{\sqrt{2}N\ms} + \cdots .
\label{nsr}\end{equation}
Combining this expansion with the energy expression (\ref{rel}),
we derive 
\begin{equation}
\me = \ms + \frac{3\sqrt{2}N}{8} - \frac{11N^2}{64\ms}
+ \frac{3\sqrt{2}\nu^2}{4N} + \cdots,
\end{equation}
where the $\ms \nu^2$ term has been canceled out.
This expression yields a non-regular $\la$ expansion
\begin{equation}
E = S + \frac{3\sqrt{2}\sqrt{\la}N}{8} - \frac{11\la N^2}{64S}
+ \frac{3\sqrt{2}J^2}{4\sqrt{\la}N} + \cdots, \hspace{1cm}
\frac{S}{\sqrt{\la}} \gg 1 \gg \frac{J}{\sqrt{\la}}.
\label{rae}\end{equation}
Here from (\ref{rat}) $N$ is much larger than $M$, 
while $N$ is the same order as $M$
in (\ref{reg}) and (\ref{jse}) for the previous cases. 
In order to trade large integer $N$ for small finite interger $M$ we use
(\ref{rat}) and (\ref{nsr}) to have $N \approx
 (2M)^{2/3}(\sqrt{2/\la}S)^{1/3}$,
whose substitution into (\ref{rae}) for $J=0$ yields
\begin{equation}
E = S + \frac{3}{4}(2M^2\la S)^{1/3} + \mathcal{O}(S^{-1/3}).
\label{eth}\end{equation}
In \cite{ART,SFT} the energy of the circular constant-radii string
solution with the large two spins $S_1, S_2$ in $AdS_5$ was presented as
\begin{equation}
E = S + \frac{3}{4}(\la S)^{1/3}\left(2k_1^2\frac{S_1}{S_2}
\right)^{1/3} + \cdots,
\label{etk}\end{equation}
where $k_1S_1 + k_2S_2 = 0, S = S_1 + S_2$ and $S_a \gg \sqrt{\la}$.
When the two spins are equal, the two winding numbers $k_a$ satisfy 
$k_1 = -k_2$ and the regular $\la$ expansion (\ref{etk}) agrees with 
(\ref{eth}). Thus the short three-spin string with large $S$, 
when $J=0$, corresponds to the circular point-like two-spin string located
near the boundary of $AdS_5$.

For $\nu \gg 1$ the solution of (\ref{mns}) takes an expansion form
\begin{equation}
\frac{m}{n} \approx 1 - \frac{\nu}{\ms} + \left(\frac{\nu}{\ms}\right)^2
- \frac{N^2}{4\ms\nu} + \frac{5N^2}{8\ms^2} + \cdots
\label{mrn}\end{equation}
owing to $\ms \gg \nu$. The energy expression (\ref{rel}) together 
with (\ref{mrn}) yields
\begin{equation}
\me \approx \ms + \nu + \frac{N^2}{8\nu} - \frac{N^2}{8\ms} + \cdots,
\label{enn}\end{equation}
where the $\ms/\nu^2$ term has been canceled out. Thus 
restoring the $\la$-dependence we have 
\begin{equation}
E \approx S + J + \frac{\la N^2}{8J} - \frac{\la N^2}{8S} + \cdots,
\hspace{1cm} \frac{S}{\sqrt{\la}} \gg \frac{J}{\sqrt{\la}} \gg 1,
\label{ras}\end{equation}
which shows a regular $\la$ expansion. Using $N^2 \approx 4M^2S/J$
that is obtained from (\ref{rat}) and (\ref{mrn}) we trade large 
integer $N$ for small finite integer $M$ to have 
\begin{equation}
E = S + J + \frac{\la M^2S}{2J^2} - \frac{\la M^2}{2J} + \cdots.
\end{equation}
Its $\la$ dependence remains the same as (\ref{ras}), while the 
$\la^{1/2}$ dependence of the subleading term in (\ref{rae}) is changed
into the $\la^{1/3}$ dependence of the corresponding term in (\ref{eth}).

\section{Conclusion}

Analyzing the conformal gauge constraint and the closed string periodicity
conditions for both the radial and angular directions in $AdS_5$, we
have constructed a solution describing a folded three-spin string with 
two equal spins in $AdS_5$ and a spin in $S^5$. This string is
orbiting around the origin and its energy is a function of the total
$AdS_5$ spin $S$, the $S^5$ spin $J$ and the folding and winding numbers 
as a parametric solution of the system of three transcendental
equations. 

We have observed that there exist three types of string configurations 
in certain limits; a long string with a large total spin $S$,
a short string with a small $S$ located near the origin of $AdS_5$ 
and a short string with a large $S$ located near the boundary of $AdS_5$.
By means of the iterative expansion method we have extracted 
the energy-spin relations for the short strings.
For the large value of $S$ there appear two different configurations such
as the long string and the latter short string, whose energy expressions
include the linear term of $S$ irrespective of the magnitude of $J$.  
The former short string is characterized by the folding and winding 
numbers $N, M$ with same order magnitude, while the latter short string
by the small value of $M/N$.
The long string as well as the former short string has been shown
to produce the same enegy-spin relations as the two-spin $(S, J)$ string,
where a logarithmic behavior,  a Regge-type one and a BMN-type one
appear. We have demonstrated that the latter short string with $J =0$
reproduces the energy-spin relation with a term $S^{1/3}$ for the 
two-spin string with two equal $AdS_5$ spins. 
We have observed that the energy expression for the latter short string
with $S \gg \sqrt{\la} \gg J$ shows a non-regular $\la$ expansion,
while that with $S \gg J \gg \sqrt{\la}$ 
is indeed arranged to take a regular $\la$
expansion. The non-perturbative $\la$ expansion for the small $S^5$ 
spin is changed into the perturbative $\la$ expansion for the large
$S^5$ spin, which transition is also seen in the long string
as well as the former short string. Thus the spin
in $S^5$ should be large in order to have the regular
$\la$ expansion for the energy of 
folded three-spin solution, which includes
the leading linear term of $J$. It is desirable to derive these regular
$\la$ expansions of the semiclassical string energy by computing the
quantum anomalous dimension of the relevant gauge invariant operator 
including covariant derivatives from the SYM perturbation theory.

\end{document}